\documentclass{aa}

\usepackage{graphicx}
\usepackage{txfonts}
\usepackage{hyperref}
\usepackage{xcolor}

\newcommand{\lya}{Ly$\alpha$}

\newcommand{\SBcgs}{$\mathrm{erg\,s^{-1}\,cm^{-2}\,arcsec^{-2}}$}

\newcommand{\rev}[1]{{\color{black}#1}}

\begin{document}

\title{
Caught in the act: diffuse stellar and Ly$\alpha$ emission around a halo-less quasar in the assembling cluster CARLA J0800+4029 at z=2
}

\author{Sofia G. Gallego
        \inst{1,2}
        \and
        Simona Mei \inst{1,3,4}
        \and
        \rev{Emanuele Daddi} \inst{\rev{5}} \and \rev{Gaël Noirot}
        \inst{\rev{6}} \and \rev{Dominika Wylezalek} \inst{\rev{7}} \and \rev{R. Michael Rich} \inst{\rev{8}}
       }

\institute{
    Universit\'e Paris Cit\'e, CNRS(/IN2P3), Astroparticule et Cosmologie, F-75013 Paris, France\\
    \email{sofiag.gallego@gmail.com}
    \and
    Cosmoventures, Paris, France
    \and
    CNRS-UCB International Research Laboratory, Centre Pierre Binetruy, IRL2007, CPB-IN2P3, Berkeley, USA
    \and
    Jet Propulsion Laboratory and Cahill Center for Astronomy \& Astrophysics, California Institute of Technology, 4800 Oak Grove Drive, Pasadena, California 91011, USA
    \and
    Universit\'e Paris-Saclay, CEA, CNRS, AIM, 91191, Gif-sur-Yvette, France \and \rev{Space Telescope Science Institute, 3700 San Martin Drive, Baltimore, MD 21218, USA} \and \rev{Astronomisches Rechen-Institut, Zentrum für Astronomie der Universität Heidelberg, Mönchhofstr. 12--14, 69120 Heidelberg, Germany} \and \rev{UCLA Division of Astronomy, Department of Physics and Astronomy, 430 Portola Plaza, Box 951547, Los Angeles, CA 90095-1547, USA}
}

\date{}

\abstract{

Extended Ly$\alpha$ emission is commonly detected around luminous quasars at $z\sim2$--3 and traces cold gas in the circumgalactic environment of the AGN. Together with the spatial distribution of galaxies, Ly$\alpha$ probes gas and galaxy assembly in high-redshift overdensities. We use deep KCWI wide-field spectroscopy and \emph{HST}/F140W imaging to investigate the assembling cluster CARLA~J0800+4029 at $z=1.986\pm0.014$.

No robust extended Ly$\alpha$ emission is detected at the radio-loud quasar position. Instead, the dominant Ly$\alpha$ emission is offset from the AGN and coincides with a disturbed structure composed of galaxies, tidal features, and diffuse intra-halo light (IHL). The detected structure has $L_{\rm Ly\alpha}=1.42\times10^{43}$ erg s$^{-1}$ and reaches surface-brightness levels of $\sim10^{-17}$~\SBcgs. At the quasar position, a matched $3\sigma$ limit gives $L_{\rm Ly\alpha}<4.20\times10^{42}$ erg s$^{-1}$, a factor of $\sim3.4$ below the detected offset emission and among the most stringent constraints at similar redshift.

The Ly$\alpha$ emission is dominated by one extended region near systemic velocity, with FWHM $\sim300$--$500$ km s$^{-1}$. The associated tidal structure has an IHL fraction of $18.8\%$, and contains a spatially coherent absorption-like decrement at $\sim+1000$ km s$^{-1}$.

The offset and alignment between Ly$\alpha$ emission and IHL suggest that gas and stars have been redistributed together during assembly. The lack of a comparable Ly$\alpha$ halo around the quasar further suggests that gas in the immediate AGN environment is ionized, depleted, or suppressed, while the offset region remains favorable for Ly$\alpha$ emission. CARLA~J0800+4029 reveals baryon redistribution during cluster assembly, with the dominant observable Ly$\alpha$ reservoir decoupled from the central AGN and linked to ongoing tidal assembly.

}


\titlerunning{Quasar--Ly$\alpha$--IHL Link in a Disrupted Protocluster at $z=2$}
\authorrunning{S. G. Gallego et al.}

\maketitle


\section{Introduction}

Galaxy clusters are the most massive gravitationally bound structures in the Universe and assemble hierarchically through the accretion of galaxies, gas, and dark matter. Their progenitors at high redshift, commonly referred to as protoclusters, provide direct constraints on when and how baryons are redistributed during the build-up of present-day cluster cores \citep{2016A&ARv..24...14O}. Large-area surveys and targeted searches have now established that overdensities with properties consistent with massive cluster progenitors are already in place by $z \sim 2$--3 \citep{2013ApJ...769...79W, 2014ApJ...786...17W}, while recent \emph{JWST} studies are pushing candidate structures to $z \sim 6$ and above \citep{2023MNRAS.518.4755A,2023ApJ...954...31C,2023ApJ...948L..14C,2023ApJ...952...74T,2024ApJ...963....9M}.

A wide range of observational techniques is used to identify and characterize these systems. One successful approach is to search for overdensities around luminous AGN, particularly radio-loud sources, where the AGN acts as a beacon for dense environments \citep{2013ApJ...769...79W, 2014ApJ...786...17W}. Complementary methods include overdensity searches using photometric redshifts and stellar-mass selections in deep multiwavelength surveys \citep{2005ApJ...626...44S,2009ApJ...694.1517D,2007A&A...468...33E}, as well as searches for line emitters, dusty star-forming concentrations, and other tracers of actively assembling environments.

Beyond their galaxy populations, assembling clusters host diffuse baryonic components whose spatial distributions encode their dynamical history. In the local Universe, tidal stripping and mergers produce intracluster light (ICL), while at earlier times this component is often referred to as intra-halo light (IHL), reflecting the fact that the system may not yet be virialized. Simulations predict that a substantial fraction of stellar mass may already be in diffuse form by $z \sim 2$, driven by frequent mergers and tidal interactions \citep{2017MNRAS.470.4186B,2018MNRAS.475..648P}. Observationally, deep \emph{HST} imaging of confirmed overdensities has begun to reveal low-surface-brightness tidal features and diffuse stellar components consistent with early-stage IHL formation \citep{2018MNRAS.474..917M,2023MNRAS.523...91W}. Establishing when this component appears and how it relates to the gaseous reservoirs of protoclusters remains an open problem.

Gas assembly and processing in overdense environments can be traced by extended Ly$\alpha$ emission. Integral-field spectrographs such as MUSE and KCWI have revealed giant Ly$\alpha$ halos around quasars and star-forming galaxies at $z \sim 2$--3, often extending to scales of $\sim 100$~kpc \citep{2016ApJ...831...39B,2019MNRAS.482.3162A}. The physical origin of this emission likely involves a combination of resonant scattering, photoionization by galaxies or AGN, and cooling or shock-heated gas.

However, the presence and morphology of Ly$\alpha$ halos vary strongly from system to system. Not all luminous quasars show extended Ly$\alpha$ emission, and suppression has been attributed to feedback, radiative-transfer effects, obscuration, or environmental and geometric factors \citep{2020A&A...635A.157T,2023A&A...680A..70W}. In particular, some radio-loud AGN appear to lack the giant Ly$\alpha$ halos commonly associated with quasars of similar luminosity, suggesting that local conditions can strongly modulate the visibility of the cold gas reservoir.

The connection between diffuse stellar light and Ly$\alpha$-emitting gas is still poorly constrained observationally. IHL traces the redistribution of stars through mergers and tidal stripping, whereas Ly$\alpha$ traces cold gas shaped by inflows, outflows, shocks, and radiative transfer in the circumgalactic and intragroup medium. A spatial correspondence or clear decoupling between these components would provide direct constraints on baryon cycling during the assembly of cluster cores at $z \sim 2$, when both merger rates and gas accretion are expected to be high.

In this work, we investigate CARLA~J0800+4029, \rev{an overdensity at $z = 1.986 \pm 0.014$ identified by the Clusters Around Radio-Loud AGN (CARLA) survey and spectroscopically confirmed with \emph{HST}/WFC3 G141 observations \citep{2013ApJ...769...79W,2014ApJ...786...17W,2016ApJ...830...90N,2018ApJ...859...38N}.} \emph{HST-WFC3IR}/F140W imaging reveals a strongly disturbed core with tidal features indicative of active assembly \citep{2023A&A...670A..58M}. We combine these data with deep \emph{Keck}/KCWI integral-field spectroscopy to search for extended Ly$\alpha$ emission and to test whether diffuse stellar and gaseous components are spatially linked.

We report a striking spatial decoupling: in the current analysis, the KCWI data recover no comparably robust extended Ly$\alpha$ signal at the quasar position, whereas the detected extended emission is dominated by one main region offset from the AGN and spatially coincident with disrupted stellar continuum and IHL in the \emph{HST} imaging. \rev{We compare this region with other diffuse structures in the field selected using the same procedure to assess the Ly$\alpha$--IHL association.} We further identify a spatially coherent absorption-like feature within the brightest Ly$\alpha$ region, indicating the presence of a cold-gas structure plausibly embedded in the tidal debris.

Taken together, these observations provide a snapshot of baryon redistribution in an assembling cluster core at $z \sim 2$, where tidal interactions appear to redistribute stars and cold gas on large scales, while Ly$\alpha$ emission is not strongly detected at the active nucleus itself. CARLA~J0800+4029 thus offers a rare view of the processes governing the emergence of diffuse baryonic components during the early stages of cluster assembly. \rev{Throughout this paper, we describe J0800+4029 as an assembling cluster and retain the term overdensity when referring to its observational selection.}

The paper is organized as follows: Section~\ref{sec:data} describes the data, Section~\ref{sec:meth} the methods, Section~\ref{sec:results} the results, Section~\ref{sec:discussion} the discussion, and Section~\ref{sec:summary} the summary. We adopt a flat $\Lambda$CDM cosmology with $\Omega_{\rm m}=0.308$, $\Omega_\Lambda=0.691$, and $H_0=67.8$~km~s$^{-1}$~Mpc$^{-1}$ \citep{2016A&A...594A..13P}. Magnitudes are in the AB system.

\section{The Data}\label{sec:data}

\subsection{CARLA: Spitzer, optical, and infrared view of J0800+4029}

CARLA~J0800+4029 is part of the Clusters Around RLAGN (CARLA) survey \citep{2013ApJ...769...79W,2014ApJ...786...17W}, which targeted 420 radio-loud AGN fields with the {\it Spitzer Space Telescope} using IRAC 3.6$\mu$m (IRAC1) and 4.5$\mu$m (IRAC2) channels (PI: D.~Stern). The survey identified galaxy (proto)clusters at $z>1.3$ through the IRAC color cut $(IRAC1-IRAC2)>-0.1$, a reliable tracer of galaxies at $z>1.3$, with the survey sensitivity favoring the brighter and typically more massive population. Roughly half the CARLA fields showed overdensities above $2\sigma$ compared to the SpUDS control field \citep{2013ApJS..206...10G}, validating the method as a cluster finder.

The twenty richest CARLA overdensities were followed up with a 40-orbit {\it HST} program using WFC3 {\it G141} grism spectroscopy and F140W ($H_{140}$) imaging (PI: D.~Stern). Sixteen fields, including J0800+4029 at $z = 1.986 \pm 0.014$, were spectroscopically confirmed as overdensities \citep{2018ApJ...859...38N}. J0800+4029, in particular, was classified as a strong cluster candidate, centered on a luminous radio-loud AGN.

Complementary $i$-band imaging was obtained with the {\it WHT}/ACAM \citep{2015MNRAS.452.2318C}, providing 6600s of exposure under 0.9\arcsec\ seeing. The data reached a $5\sigma$ limit of 25.16 mag and fully covered the {\it HST} footprint.

Stellar population analyses comparing J0800+4029 to another CARLA system at similar redshift (J2039--2514) revealed clear differences: while J2039--2514 already shows a red sequence, J0800+4029 is dominated by star-forming galaxies \citep{2016ApJ...830...90N}, suggesting an earlier dynamical stage.

Recent work \citep{2023A&A...670A..58M} \rev{identified 19 photometrically selected candidate members} within 0.5\arcmin\ of the overdensity peak. They measured a high overdensity significance ($S/N_c = 17$) with $\sim10\%$ contamination. The core stellar mass was found to be $\log(M_*/M_\odot) = 12.3$, implying a halo mass $\log(M_{\rm halo}/M_\odot) \sim 14.3$. The population remains dominated by late-type galaxies, with a merger fraction an order of magnitude higher than the field, indicating ongoing transformations.

Despite its large stellar mass, the system hosts a passive galaxy fraction more typical of groups than clusters, consistent with a young stage. The high merger rate and low-surface-brightness tidal features visible in $H_{140}$ images directly point to stellar redistribution and early intra-halo light (IHL) formation.

\begin{figure*}
\includegraphics[trim={0cm 0cm 0cm 0cm},clip,width=2\columnwidth]{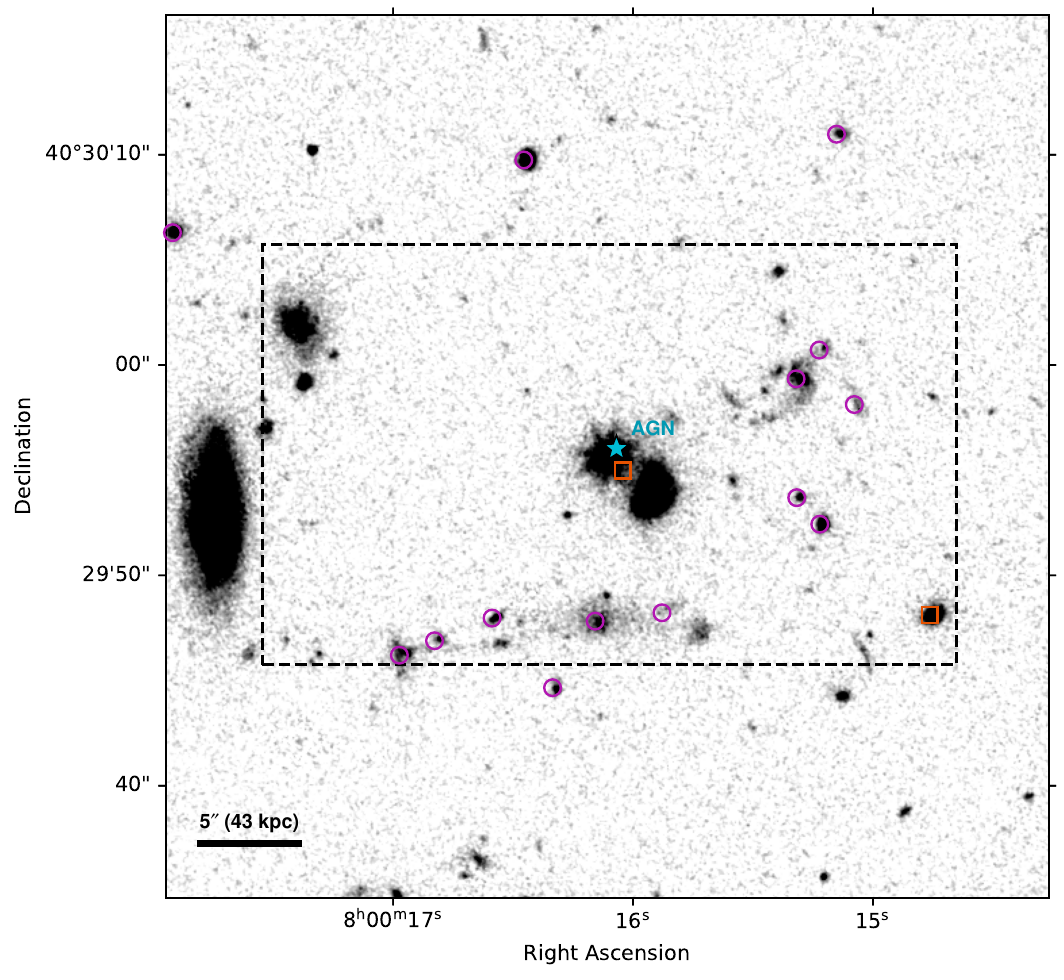}
\caption{Core of CARLA~J0800+4029 in {\it HST} $H_{140}$. \rev{The cyan star marks the radio-loud AGN B3~0756+406; orange squares show two sources with catalogued redshifts in $1.9\leq z\leq2.1$, and magenta circles show the photometrically selected candidates of \citet{2023A&A...670A..58M}.} The central AGN is near a merging pair at $\sim 3$\arcsec\ separation. Additional late-type galaxies are linked by faint tidal features to the west and south. The dashed rectangle marks the KCWI field of view ($20''\times33''$). \rev{The $5''$ scale bar corresponds to 43~kpc at $z=1.986$.}}
\label{fig:mergers}
\end{figure*}

Notably, an interacting galaxy pair lies $\sim 3$\arcsec\ from the AGN. The {\it HST} image also reveals groupings of late-type galaxies connected by faint tidal features and stellar streams, signaling merger-driven cluster growth (Figure~\ref{fig:mergers}).

\subsection{KCWI Observations}

We obtained integral-field spectroscopy of J0800+4029 with the Keck Cosmic Web Imager (KCWI) on Keck II. Observations used the medium-blue setup, covering 3500--5500~\AA\ to capture Ly$\alpha$ at the cluster redshift. The $20''\times33''$ field of view was sampled at 0.29\arcsec\ pixel$^{-1}$ with spectral resolution $R\sim4000$ ($\sim 70$ km\,s$^{-1}$).

The dataset consists of 3600s on source, split into three 1200s exposures with dithers. Pointing was centered on the AGN to maximize sensitivity to extended Ly$\alpha$.

Data were reduced with the KCWI pipeline: bias subtraction, flat-fielding, wavelength calibration, illumination correction, sky subtraction, and cosmic ray removal (via LA Cosmic). Exposures were registered and coadded, producing a final cube with 0.3\arcsec\ pixels and $\sim1.0$~\AA\ sampling.

To enhance faint diffuse emission, we applied additional background treatment to suppress residual large-scale structure. \rev{Non-finite and zero-valued voxels in the working cube were excluded from the adaptive-kernel detection and subsequent measurements.}

\section{Methods} \label{sec:meth}

\subsection{Diffuse \lya\ Detection}

Faint extended Ly$\alpha$ emission was searched for using an Adaptive Kernel Smoothing (AKS) approach \citep[e.g.,][]{2019NatAs...3..822M,2020ApJ...894....3O,2021A&A...649A..78D,2025A&A...698A.243G}. In practice, the search uses a conservative range of spatial smoothing kernels, from the native KCWI resolution up to scales of order $\sim2''$, together with a minimum connected-volume requirement designed to suppress spurious detections. The current implementation adopts a minimum size of 500 connected voxels.

Detected voxels were grouped into candidate three-dimensional components and then inspected in both the raw and smoothed data. In the present dataset, the recovered Ly$\alpha$ signal is dominated by one main extended region, although the exact decomposition into subcomponents depends on the adopted 3D labeling scheme. We therefore use the AKS products primarily to define robust candidate emission regions and to guide subsequent moment-map and spectral extractions.

\subsection{Upper Limit on a Quasar-centered \lya\ Halo}
\label{subsec:qso_limit}

An upper limit on extended Ly$\alpha$ emission at the quasar position is estimated from the local noise properties of the raw cube. The sensitivity is computed by propagating the variance within a standard circular aperture centered on the AGN and over a 7-channel spectral window around systemic Ly$\alpha$. Matched blank-aperture measurements in nearby unmasked regions are used as an empirical cross-check on the variance-based estimate.

For comparison with published halo luminosities, this local noise estimate is also converted into an upper limit for a hypothetical quasar-centered halo matched in projected area and spectral width to the detected offset/tidal Ly$\alpha$ complex. This provides a like-for-like constraint on the luminosity of any halo that would have been present at the AGN position with morphology and spectral extent comparable to the detected offset structure.

The central source is masked in the working cube to avoid contamination from the bright quasar emission, but this masked region is small compared to the spatial extent expected for a giant Ly$\alpha$ halo. The quasar position is therefore treated as a non-detection for the purposes of the present analysis, while the halo-scale upper limit is set by the surrounding local noise.

\subsection{Absorption Feature Extraction}
\label{subsec:absorption_method}

Within the brightest extended Ly$\alpha$ region, we identified a localized flux decrement adjacent to the diffuse emission profile. The candidate feature was first recognized in the smoothed KCWI products and then verified directly in the raw cube to ensure that it was not introduced by the adaptive-kernel procedure.

To characterize the feature, we combined three complementary diagnostics. First, we constructed a surface-brightness map integrated over the spectral channels spanning the minimum of the decrement, in order to test whether the deficit is spatially coherent across multiple contiguous spaxels. Second, we extracted spectra in concentric apertures centered on the decrement to test whether the feature persists with aperture size and to trace how its depth and centroid vary with radius. Third, we compared the morphology of the decrement with the \emph{HST}/F140W image and with the extended Ly$\alpha$ emission contours in order to assess its spatial relation to the tidal stellar structure.

Because the feature is superposed on broad diffuse Ly$\alpha$ emission, we do not model it as an isolated absorption line in the classical sense. Instead, we treat it conservatively as an absorption-like decrement or profile dip embedded within the larger Ly$\alpha$ halo. Its significance is therefore evaluated from its spatial coherence, its persistence across independent spectral extractions, and its consistent velocity offset relative to the systemic redshift, rather than from a single-pixel or single-spectrum detection metric.

\subsection{Moment Maps and Spectral Extraction}
\label{subsec:moments_method}

Moment maps were constructed from the detected Ly$\alpha$ spectral support in order to characterize morphology and kinematics. Surface-brightness maps trace the integrated emission within the detected region, while velocity offsets are reported relative to the systemic redshift adopted for the protocluster, $z=1.986$. In practice, the moment products are used as descriptive diagnostics of the morphology and kinematics of the detected Ly$\alpha$ structure rather than as a full dynamical model.

Regions of interest were defined from the dominant surface-brightness morphology and from the corresponding spectral support in the cube. Spectra were extracted for each candidate component and for the main tidal/Ly$\alpha$ region. Because the absorption-like feature  appears as a localized negative decrement adjacent to the extended Ly$\alpha$ halo, it does not contribute to the positive-emission spectral support used for the main Ly$\alpha$ measurements. It is therefore characterized separately from the global halo emission.

\subsection{Intra-halo Light Extraction}
\label{subsec:ihl}

Diffuse intra-halo light is measured from the background-subtracted {\it HST} F140W image. Rather than a parametric light-profile decomposition, we adopt a non-parametric, segmentation-based approach because the system is highly disturbed, with irregular morphologies and multiple tidal components.

\rev{We use the segmentation map associated with the reduced \emph{HST} image to identify compact sources. After subtracting the supplied F140W background model, we estimate the residual local background with \texttt{photutils.Background2D}, using 30-pixel boxes, a $3\times3$ median filter, and a median estimator.} The diffuse residual image is then smoothed to enhance low-surface-brightness structure, and connected regions are measured within a circular aperture of radius $17\arcsec$ centered on J0800+4029. In the fiducial setup, diffuse emission is detected at $2.5\sigma$ with a minimum of 5 connected pixels, the detection mask is smoothed with a Gaussian kernel of $\sigma=6$ pixels, contours are drawn at a fixed level of 0.2, and enclosed holes are filled to define the final connected diffuse regions. \rev{Hole filling closes gaps produced by masked compact-source pixels inside a connected diffuse contour; the compact-source flux remains assigned to the galaxy component.} Fluxes are estimated for the total light and for the galaxy-associated component, and the IHL fraction is defined as
\[
f_{\rm IHL} = \frac{F_{\rm IHL}}{F_{\rm total}},
\]
with $F_{\rm IHL}=F_{\rm total}-F_{\rm gal}$.

\rev{Projected associations are defined from a positional list assembled from a NED query. We select sources identified as the photometric candidates of \citet{2023A&A...670A..58M}, together with sources having catalogued redshifts $1.9\leq z\leq2.1$. Each position is represented by a $1.5\arcsec$-radius aperture followed by a two-pixel dilation. Diffuse regions with at least 5\% overlap with the combined mask form the candidate-associated sample; the remaining regions form the comparison sample. This mask is used only to classify diffuse regions by projected association, so Table~\ref{tab:ihl_regions} reports region-level measurements rather than object-by-object membership.}

\rev{At $z=1.986$, H$\beta$ and [O\,III] $\lambda\lambda4959,5007$ fall within F140W. We cross-match the diffuse regions with the published G141 emission-line catalog of \citet{2018ApJ...859...38N} to evaluate contamination from known compact line emitters.}

To assess robustness, the analysis includes four complementary validation tests. First, the statistical uncertainty on $f_{\rm IHL}$ is estimated by bootstrap resampling of the galaxy-assigned and diffuse-assigned pixels within each region. Second, the science-region mask is translated to blank positions within the same cutout to construct an empirical null distribution for residual diffuse flux. Third, a sign-flipped residual-map test is used to check whether comparably extended negative features are present. Fourth, the measurement is repeated under controlled variations of the background mesh, detection threshold, smoothing scale, contour level, and galaxy-mask growth in order to estimate the dominant systematic sensitivity. \rev{The tested values are 20, 30, and 40 pixels for the background box; $2.0$, $2.5$, and $3.0\sigma$ for the detection threshold; 4, 6, and 8 pixels for the smoothing scale; 0.15, 0.20, and 0.25 for the contour level; and 0, 1, and 2 pixels for galaxy-mask dilation. Bootstrap and blank-aperture tests use 2000 and 500 realizations, respectively.} In practice, the fiducial IHL fractions are most useful for relative comparison within the field, while the absolute normalization remains sensitive to the adopted galaxy/IHL separation.

\section{Results}
\label{sec:results}

\begin{figure*}
\includegraphics[trim={0cm 0cm .25cm 0cm},clip,width=2\columnwidth]{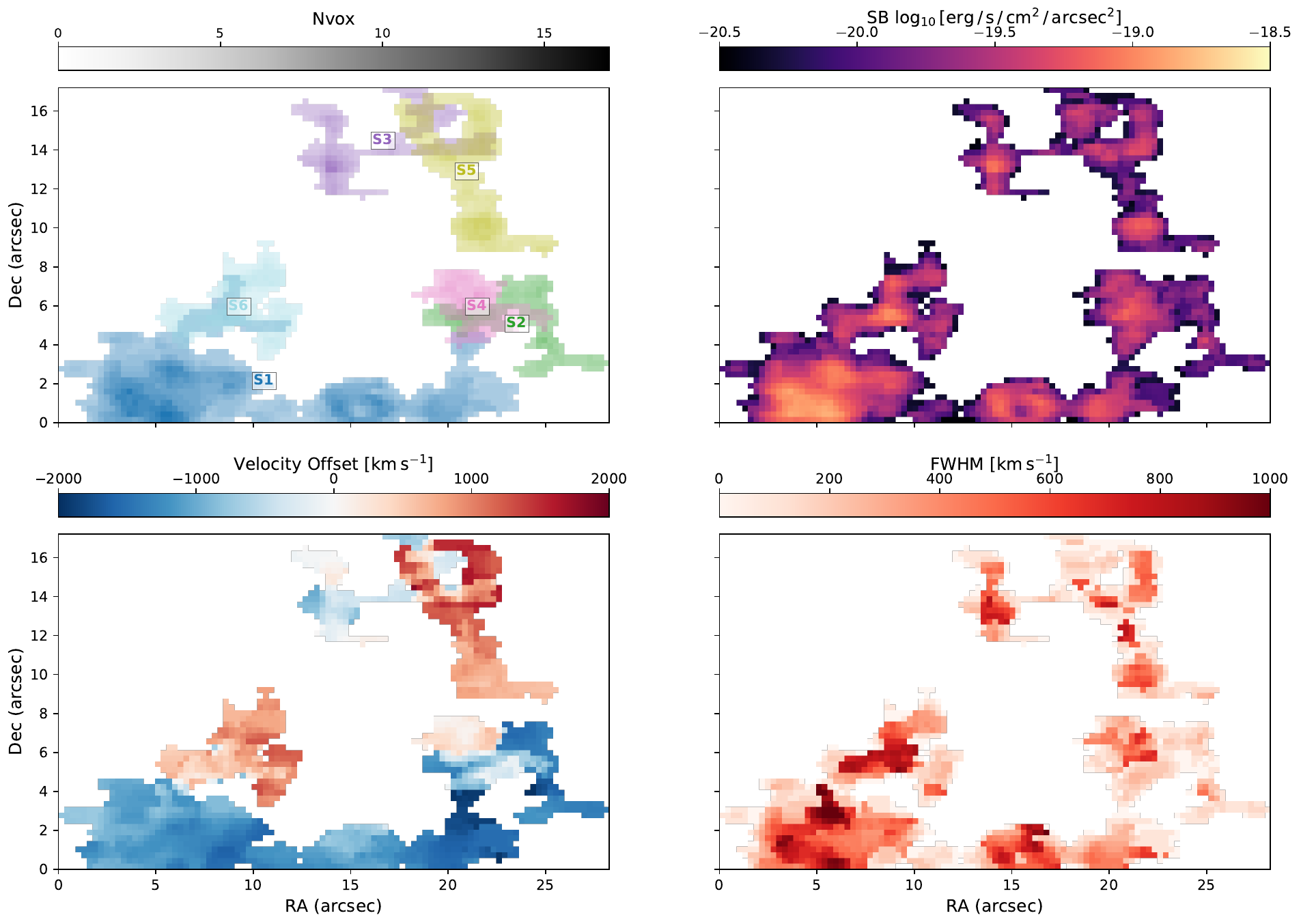}
\caption{Combined moment maps of the detected Ly$\alpha$ field. \rev{Labels S1--S6 identify the six connected three-dimensional AKS components; S1 is the dominant extended component.}}
\label{fig:moments_combined}
\end{figure*}

\begin{figure}
\includegraphics[trim={0cm 0cm .3cm 0cm},clip,width=\columnwidth]{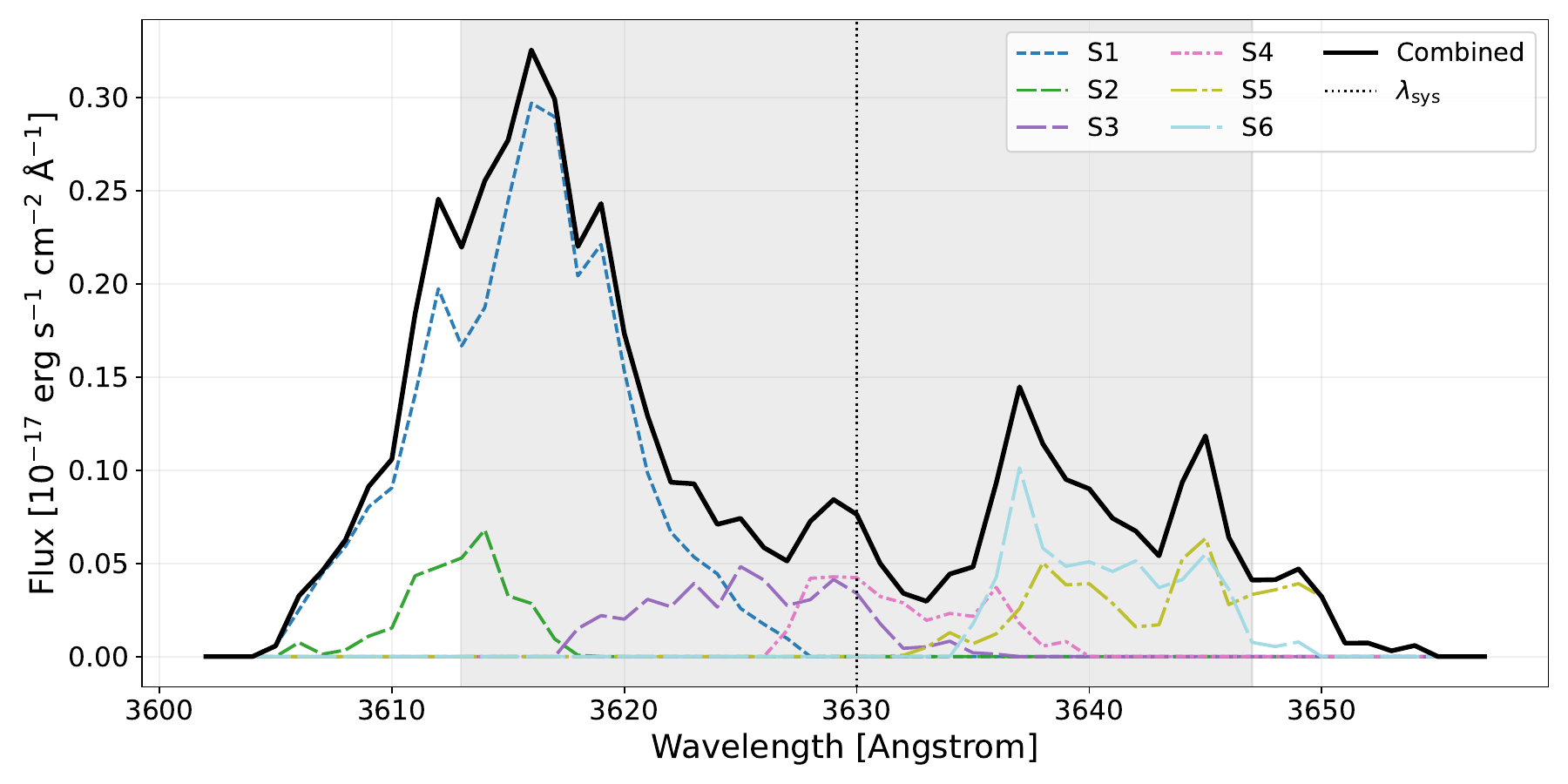}
\caption{Spectra of the detected Ly$\alpha$ components. The vertical line and shaded region mark the systemic Ly$\alpha$ wavelength and the corresponding redshift uncertainty. \rev{The labels S1--S6 correspond to the connected components shown in Fig.~\ref{fig:moments_combined}.}}
\label{fig:spectra_structures}
\end{figure}

In the current KCWI analysis, no comparably robust extended Ly$\alpha$ signal is recovered at the position of the quasar itself, despite its strong radio output. Instead, the detected Ly$\alpha$ emission is dominated by one main spatially extended region offset from the quasar. This region spatially coincides with diffuse stellar continuum and intra-halo light in the \emph{HST}/F140W imaging. Additional lower-significance or spectrally distinct components are present in the 3D AKS decomposition, but they do not dominate the morphology of the diffuse field.
For a standard $1\arcsec$-radius aperture at the quasar position and a 7-channel window around systemic Ly$\alpha$, the raw-cube noise implies a conservative $1\sigma$ surface-brightness sensitivity of $2.72\times10^{-18}$~\SBcgs\ and a $3\sigma$ limit of $8.16\times10^{-18}$~\SBcgs. If we assume a quasar-centered halo with the same projected area as the detected offset/tidal complex ($149.8$ arcsec$^2$) and the same full spectral support (50 channels), the equivalent $3\sigma$ total-halo luminosity limit is $4.20\times10^{42}$ erg~s$^{-1}$.

\subsection{Ly$\alpha$ moment maps}
\label{subsec:moments}

Figure~\ref{fig:moments_combined} presents the Ly$\alpha$ moment maps for the detected field, while Fig.~\ref{fig:spectra_structures} shows the spectra of the AKS-defined Ly$\alpha$ components. The Ly$\alpha$ signal is clearly dominated by one extended structure, which contains most of the high-signal-to-noise voxels and therefore drives the robust kinematic properties of the system.

In the surface-brightness map, the emitting structure appears as a coherent region of enhanced Ly$\alpha$ emission clearly offset from the quasar position. The voxel-count map confirms that the brightest part of the region is well sampled spectrally, ensuring that the derived kinematics are anchored by the most significant pixels.

The velocity-offset map shows that the brightest part of the Ly$\alpha$ structure lies close to the adopted systemic velocity. Across the high-voxel region, velocity offsets remain modest and do not form a simple large-scale monotonic gradient. Apparent excursions to larger positive or negative velocities are confined mainly to lower-surface-brightness pixels and should therefore be treated with caution.

The FWHM map reveals moderate line broadening within the dominant Ly$\alpha$ structure. In the brightest and best-sampled regions, the FWHM typically lies in the range $\sim300$--$500$ km~s$^{-1}$. Broader values are present in lower-surface-brightness or less securely sampled pixels, indicating disturbed kinematics but not requiring a single uniform dispersion across the entire structure.

Overall, the moment maps show that the Ly$\alpha$ emission in J0800+4029 is dominated by one main extended region that is kinematically close to systemic and characterized by disturbed but spatially coherent line broadening.
Summing the six AKS-defined Ly$\alpha$ components in the raw cube gives a total detected flux of $4.99\times10^{-16}$ erg~s$^{-1}$~cm$^{-2}$, corresponding to a total Ly$\alpha$ luminosity of $1.42\times10^{43}$ erg~s$^{-1}$ for the offset/tidal complex.

\subsection{Intracluster Light Fraction in the Tidal Region}

The dominant Ly$\alpha$ emission region is spatially coincident with the main tidal feature visible in the \emph{HST}/F140W image. No similarly clear extended Ly$\alpha$ counterpart is identified around the other diffuse structures visible in the field.

\begin{figure}
\includegraphics[trim={0cm 0cm .3cm 0cm},clip,width=\columnwidth]{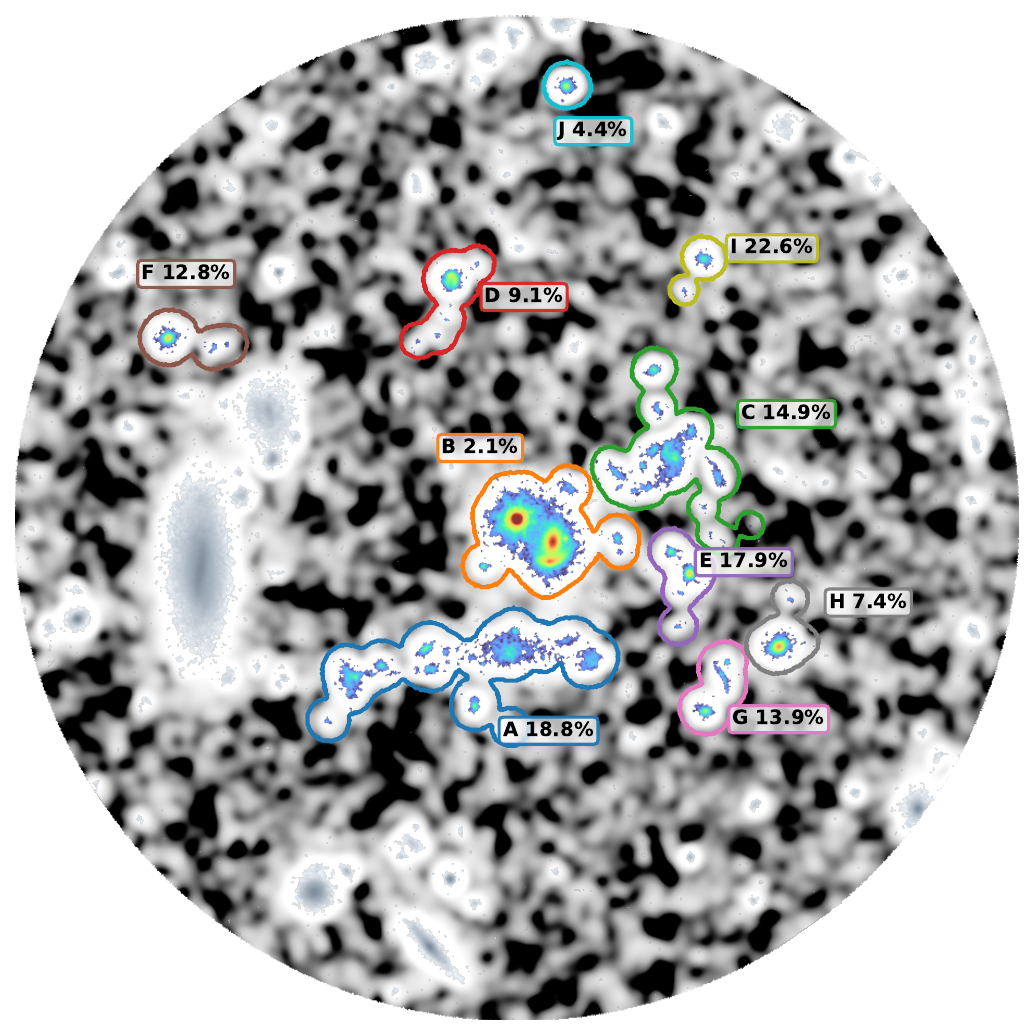}
\caption{IHL-selected regions in the \emph{HST} F140W field. Galaxies are identified and masked using a segmentation map; the galaxy-masked image is then background-subtracted and smoothed to enhance the low-S/N diffuse signal. Colored contours delineate the selected diffuse regions summarized in Table~\ref{tab:ihl_regions}. Galaxies overlapping the selected IHL regions are shown in color, while the remaining galaxies are shown with a muted blue-gray scale for visual context. Labels indicate the region ID and measured IHL fraction.}
\label{fig:HST-IHL}
\end{figure}

Figure~\ref{fig:HST-IHL} shows the 10 diffuse regions that satisfy the adopted IHL selection, labeled A through J in order of decreasing projected area, and Table~\ref{tab:ihl_regions} lists their measured IHL fractions and projected areas. Within this displayed sample, the inferred fractions span $2.1\%$--$22.6\%$. The Ly$\alpha$-associated tidal structure corresponds to region A, for which we measure $f_{\rm IHL}=18.8 \pm 0.5\%$. This places the tidal/Ly$\alpha$ structure among the highest-IHL regions in the selected sample. \rev{The 18.8\% value characterizes region A; measuring a cluster-wide IHL fraction requires integrating the complete candidate-galaxy population within a common aperture.}

\rev{No catalogued $z\sim2$ emission-line source lies within region A. The nearest source with detected [O\,III] is $1.9\arcsec$ outside its boundary; the [O\,III] doublet contributes approximately 4\% of its integrated F140W flux. A known compact [O\,III] emitter is therefore unlikely to produce the morphology of region A, although diffuse or uncatalogued line emission may contribute. We consequently report a diffuse F140W-light fraction.} The Ly$\alpha$-associated region is spatially extended, connected to disrupted stellar continuum, and morphologically distinct from the more compact or less obviously tidal diffuse structures elsewhere in the field. The bootstrap uncertainty on the science region is small compared to the object-to-object spread across the displayed sample.

The diffuse signal itself appears robust. Translating the science-region mask to blank positions in the same cutout yields an empirical null probability of $p \simeq 0.002$ for obtaining an equal or larger diffuse flux per pixel from residual background structure alone. A sign-flipped residual-map test likewise produces no negative feature with an area comparable to the science region. By contrast, a matched \rev{comparison sample below the candidate-overlap threshold} shows that the science region is elevated but not uniquely extreme in $f_{\rm IHL}$ when compared against regions of similar size, projected radius, and mean continuum brightness.

These tests support the interpretation that the diffuse structure associated with the offset Ly$\alpha$ emission is astrophysically real, while also showing that the exact numerical value of $f_{\rm IHL}$ remains method-dependent. The most robust conclusion is therefore not that the system has a uniquely extreme IHL fraction, but that the offset Ly$\alpha$ halo is associated with one of the most diffuse and tidally disturbed stellar structures in the protocluster core.

\begin{table}
\centering
\caption{Diffuse regions displayed in Fig.~\ref{fig:HST-IHL}, ordered by decreasing projected area. Quoted uncertainties are bootstrap 68\% intervals. Region A is the Ly$\alpha$-associated tidal structure.}
\label{tab:ihl_regions}
\small
\begin{tabular}{lcc}
\hline
Region & $f_{\rm IHL}$ (\%) & Area (arcsec$^2$) \\
\hline
A & $18.8 \pm 0.5$ & 23.76 \\
B & $2.1 \pm 0.2$  & 17.16 \\
C & $14.9 \pm 0.7$ & 16.30 \\
D & $9.1 \pm 1.4$  & 6.19 \\
E & $17.9 \pm 1.4$ & 5.25 \\
F & $12.8 \pm 1.4$ & 4.99 \\
G & $13.9 \pm 2.0$ & 4.82 \\
H & $7.4 \pm 0.8$  & 4.71 \\
I & $22.6 \pm 2.5$ & 2.52 \\
J & $4.4 \pm 2.1$  & 1.94 \\
\hline
\end{tabular}
\end{table}

\subsection{Absorption Feature Detection}
\label{subsec:absorption}

Within the brightest region of diffuse Ly$\alpha$ emission, we detect a localized absorption-like feature superposed on the emission profile. The feature is spatially resolved and coincident with the tidal/IHL structure identified in the \emph{HST} imaging, while being clearly offset from both the quasar position and the radio source.

\begin{figure*}
  \centering
  \includegraphics[width=\textwidth]{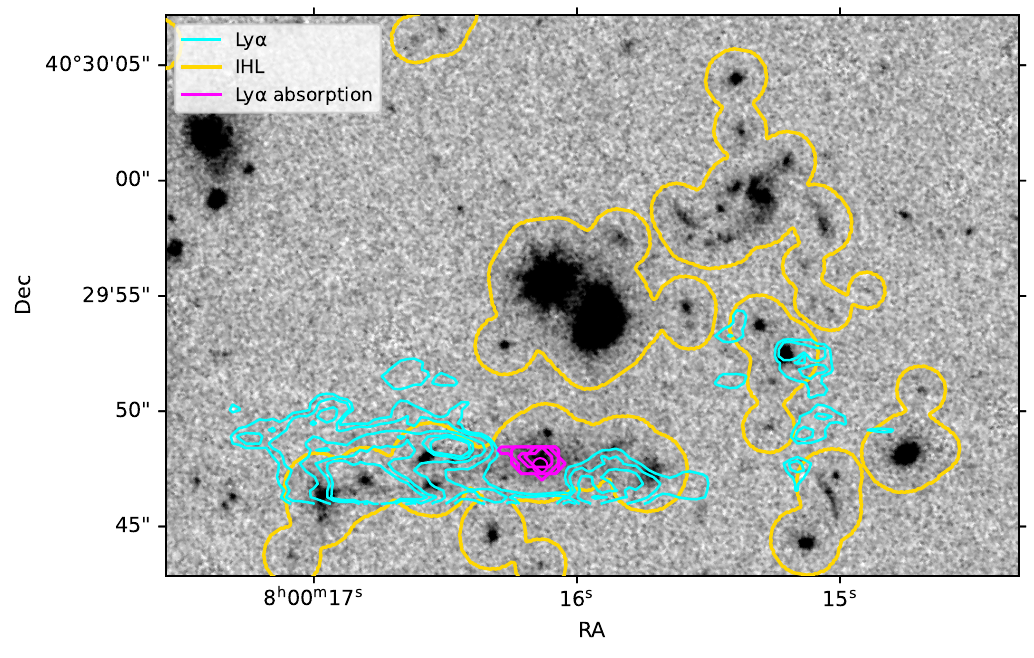}
  \caption{
  Composite view linking diffuse stellar and gaseous components in CARLA J0800+4029.
  The HST/F140W image traces tidal features and intra-halo light (IHL), while Ly$\alpha$
  emission contours from KCWI are overlaid. The location of the absorption-like Ly$\alpha$
  decrement is highlighted, showing its spatial coincidence with the tidal/IHL structure
  and its clear offset from the central quasar.}
  \label{fig:ihl_lya_absorption}
\end{figure*}

Figure~\ref{fig:ihl_lya_absorption} illustrates the spatial association between the tidal stellar continuum, the extended Ly$\alpha$ emission, and the localized absorption-like feature, highlighting their joint offset from the active nucleus.

\begin{figure}
\centering
\includegraphics[width=\columnwidth]{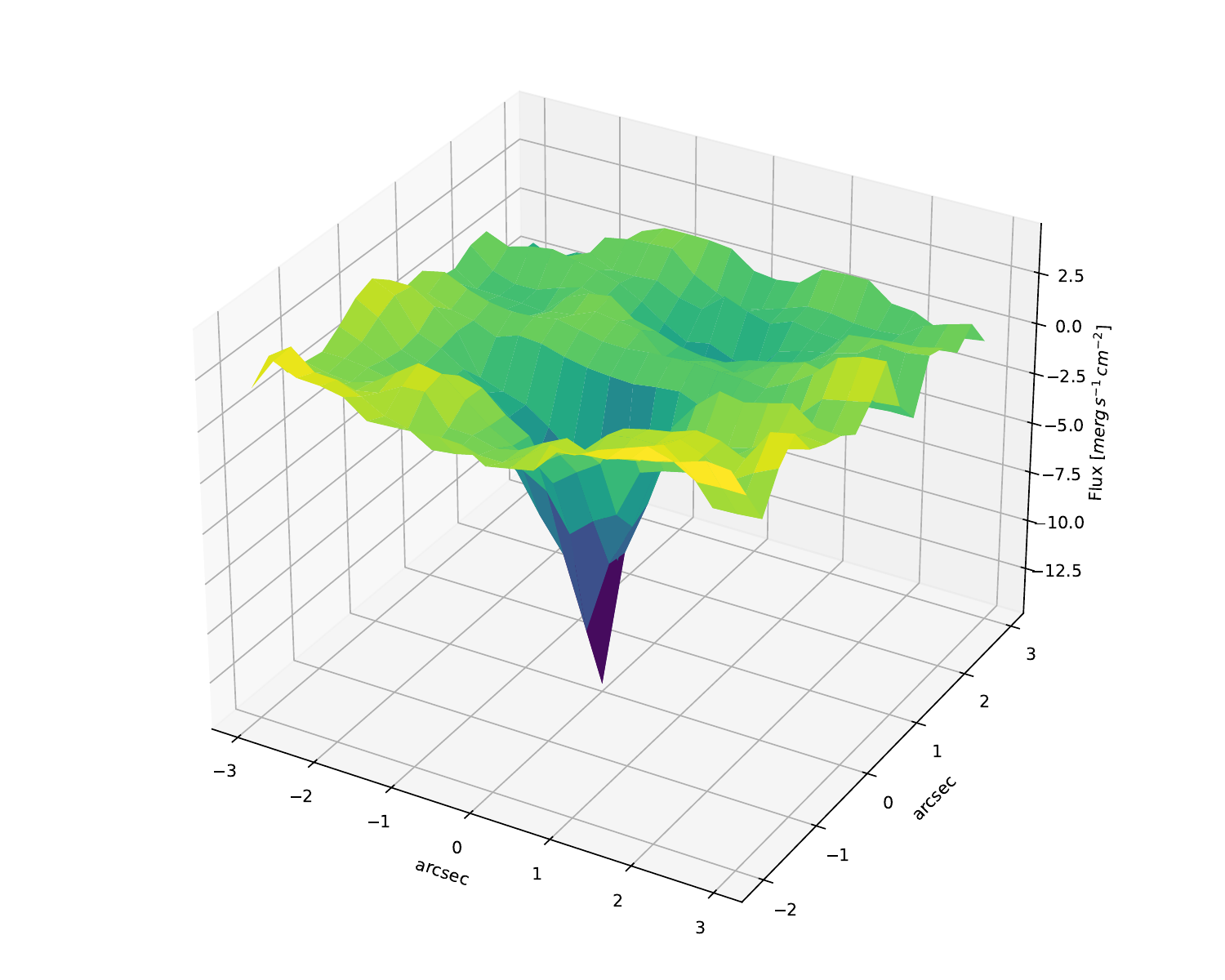}
\caption{Integrated Ly$\alpha$ surface-brightness map highlighting the absorption-like feature. The deficit is spatially coherent and embedded within the extended Ly$\alpha$ structure associated with the tidal/IHL region.}
\label{fig:abs_surface}
\end{figure}

Figure~\ref{fig:abs_surface} presents the Ly$\alpha$ surface-brightness depression integrated over the spectral channels corresponding to the absorption minimum. The deficit extends coherently over several contiguous spaxels and does not correlate with regions of low exposure, elevated noise, or obvious instrumental artifacts.

\begin{figure}
\centering
\includegraphics[width=\columnwidth]{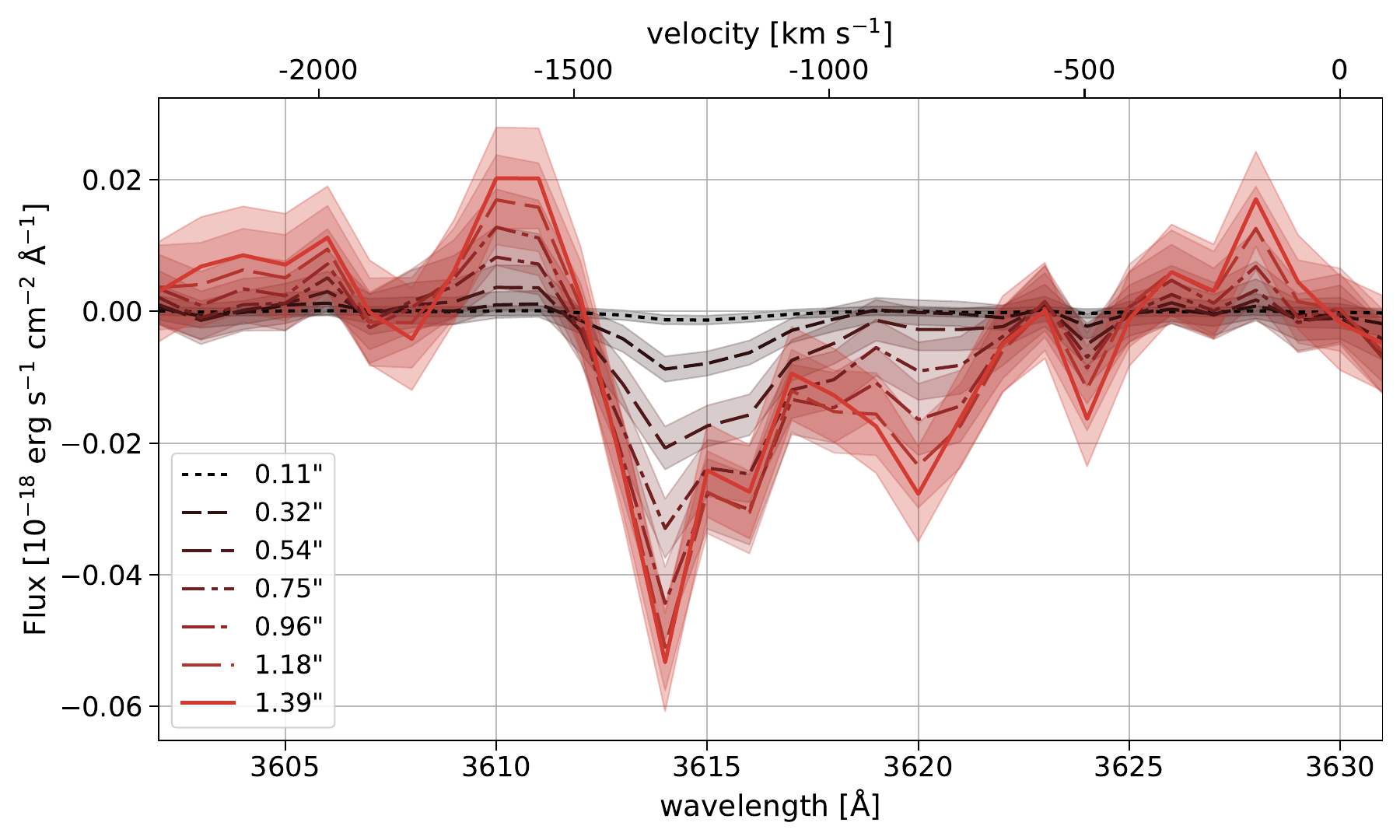}
\caption{Ly$\alpha$ spectra extracted in radial bins centered on the absorption-like feature. A coherent decrement is visible at $\sim +1000$ km~s$^{-1}$ relative to systemic, persisting across multiple radial apertures.}
\label{fig:abs_spectra_radial}
\end{figure}

Ly$\alpha$ spectra extracted in concentric radial bins centered on the feature are shown in Fig.~\ref{fig:abs_spectra_radial}. A persistent decrement is detected at a velocity offset of $\sim +1000$ km~s$^{-1}$ relative to the systemic redshift. Both the centroid and depth of the feature vary smoothly with radius, arguing against a chance superposition of isolated noise spikes. Velocity-selected spectral extractions further show that the decrement preferentially affects the redshifted side of the Ly$\alpha$ profile.

The spatial coherence of the surface-brightness deficit, its consistent velocity offset, and its persistence across independent spectral extractions indicate that the feature is likely real. Its location within the tidal/IHL region suggests an association with cold gas embedded in the stellar debris, although inflow, complex radiative transfer, and disturbed kinematics remain viable interpretations.

\section{Discussion} \label{sec:discussion}

The results presented above reveal a clear spatial decoupling between the active nucleus, the dominant diffuse Ly$\alpha$ emission region, and the intra-halo light in CARLA~J0800+4029. Rather than a centrally concentrated gaseous halo obviously centered on the quasar, the current analysis is dominated by offset Ly$\alpha$ emission coincident with tidal stellar structures, together with a localized absorption-like feature tracing cold gas.

This configuration points to a non-trivial redistribution of baryons within the protocluster core. The lack of a comparably robust extended Ly$\alpha$ detection at the quasar position, together with offset diffuse emission, IHL, and an embedded absorption-like feature, indicates that gas and stars do not respond uniformly to the same physical drivers. Interpreting these components jointly places this system within the broader context of gas availability, feedback efficiency, radiative transfer, and tidal processing at $z\sim2$.

\subsection{Quasar Position and the Lack of Extended \lya\ Emission}

In the current analysis, the quasar position does not show an extended Ly$\alpha$ signal comparable to that recovered in the offset tidal region. This result is unusual for a radio-loud quasar at this redshift and suggests that the gas conditions near the AGN differ from those in the surrounding disturbed structure.
Using the halo-matched comparison described above, the $3\sigma$ total-halo luminosity limit at the AGN position is about a third of the total detected Ly$\alpha$ luminosity of the offset/tidal complex. In that sense, a quasar-centered halo comparable to the observed offset structure would still have been strongly disfavored by the current noise level, even though the masked central region prevents a direct flux measurement at the nucleus itself.

One plausible interpretation is that feedback from the AGN, whether mechanical or radiative, reduces the visibility of Ly$\alpha$ in its immediate environment. However, dust attenuation, resonant radiative-transfer effects, viewing geometry, and local gas distribution may all affect whether extended Ly$\alpha$ is observable around the nucleus. We therefore interpret the quasar non-detection cautiously: it is a robust feature of the current analysis, but not by itself a unique signature of feedback-driven suppression.

\subsection{Diffuse \lya\ Emission and the Intra-halo Light}

The offset Ly$\alpha$ emission spatially coincident with the intra-halo light (IHL) indicates that the dominant reservoir of observable gas is associated with the tidal structure rather than with the quasar itself. This configuration is consistent with a common origin in a recent interaction, where both stars and gas have been redistributed on comparable spatial scales. In this framework, the diffuse stellar component traces stripped stars, while the Ly$\alpha$ emission arises from gas displaced during the same event and made visible through a combination of shock heating, illumination, or resonant scattering \citep[e.g.,][]{2023A&A...677A...3C}. The morphology observed in J0800+4029, in which the main Ly$\alpha$ structure follows the tidal/IHL feature, is naturally explained in this context.

The absence of a similarly extended Ly$\alpha$ halo at the quasar position suggests that the gas conditions in the immediate AGN environment differ from those in the tidal structure. One plausible interpretation is that the Ly$\alpha$-bright region traces a recently accreted, gas-rich system that has not yet been fully mixed or processed within the central halo. Such a structure would provide a localized reservoir of relatively cold gas in the surrounding CGM, enhancing the visibility of Ly$\alpha$ emission. In contrast, gas near the quasar may be more highly ionized, depleted, or affected by radiative transfer and feedback processes, reducing the detectability of extended Ly$\alpha$ emission in that region.

Although the present data do not uniquely constrain the origin or kinematics of the gas, the combination of spatial offset, morphological coherence, and the absence of a central Ly$\alpha$ halo supports a scenario in which the observed emission traces gas recently redistributed by dynamical interactions, rather than gas in a relaxed, centrally concentrated configuration within the protocluster core.

\subsection{Absorption-like Feature and Cold Gas}

The redshifted absorption-like feature detected within the Ly$\alpha$ halo indicates that the gas is multiphase and kinematically complex. An inflowing cold-gas component is one plausible explanation, particularly given the feature's velocity offset and spatial association with the disturbed stellar debris. However, the present data do not require a unique inflow interpretation.

Alternative explanations include radiative-transfer effects within a complex Ly$\alpha$ emitting medium, patchy attenuation, or a superposition of dynamically disturbed gas components. More detailed line-profile modeling or independent constraints from non-resonant lines would be needed to distinguish among these scenarios.

\subsection{Comparison with other high-redshift halos}
\begin{figure}
\includegraphics[trim={0cm 0cm 0cm 0cm},clip,width=\columnwidth]{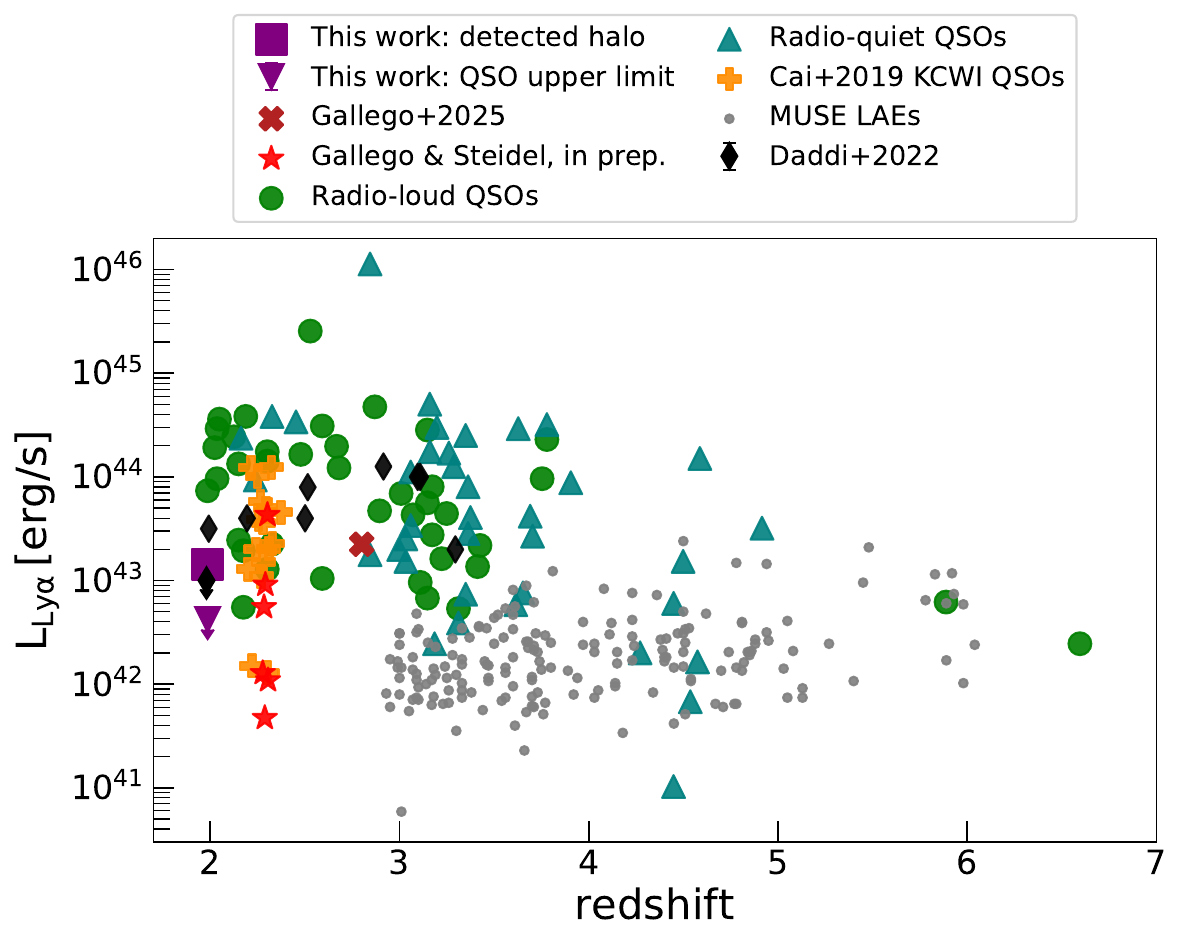}
\caption{Ly$\alpha$ halo luminosity as a function of redshift. The green circles and teal triangles correspond to Ly$\alpha$ halos detected around radio-loud and radio-quiet quasars, respectively. The orange plus symbols show the KCWI QSO halos at $z\approx2$ from \citet{2019ApJS..245...23C}. The gray dots correspond to Ly$\alpha$ halos around LAEs observed with MUSE \citep{2017A&A...608A...8L}. The black diamonds show the galaxy groups and clusters compiled by \citet{2022ApJ...926L..21D}, with the downward-pointing limit corresponding to XLSSC~122. The red stars correspond to the halo complex detected in the J0103+1316 protocluster at $z=2.3$ (Gallego \& Steidel, in preparation). The dark red cross symbol marks the halo reported in \citet{2025A&A...698A.243G} at $z=2.8$. The purple square shows the detected offset/tidal Ly$\alpha$ halo in this work, while the purple downward-pointing symbol marks the $3\sigma$ upper limit at the quasar position, assuming a halo matched in projected area and spectral width to the detected complex.}
\label{fig:lum-vs-z}
\end{figure}

The total detected offset/tidal Ly$\alpha$ emission in this work corresponds to $L_{\mathrm{Ly}\alpha}=1.42\times10^{43}$ erg s$^{-1}$ ($\log_{10}L_{\mathrm{Ly}\alpha}=43.15$), while the quasar-position halo-matched $3\sigma$ upper limit is $L_{\mathrm{Ly}\alpha}<4.20\times10^{42}$ erg s$^{-1}$ ($\log_{10}L_{\mathrm{Ly}\alpha}<42.62$). The detected offset halo therefore lies in the lower-luminosity part of the extended Ly$\alpha$ halo distribution, overlapping more naturally with group and protocluster-scale systems than with the brightest QSO-centered nebulae. In contrast, the limit at the AGN position falls below most of the QSO halos at comparable redshift, including the KCWI sample of \citet{2019ApJS..245...23C}, which reinforces the conclusion that the dominant Ly$\alpha$ emission in our system is spatially displaced from the quasar. The halo reported in \citet{2025A&A...698A.243G} at $z=2.8$, is only modestly brighter than the detected offset halo in our field, whereas the J0103+1316 protocluster halos span a broader range of luminosities at similar redshift. Together, these comparisons place our detection in the regime of moderate-luminosity extended Ly$\alpha$ structures while emphasizing the non-detection of a bright nuclear halo around the AGN itself.

\subsection{Comparison with other high-redshift protoclusters}

Extended Ly$\alpha$ halos are known to depend strongly on environment and feedback \citep[e.g.,][]{2013ApJ...770...57B,2020A&A...641A...6P}, and offset halos remain rare partly because of limited IFU coverage.

The Spiderweb protocluster at $z=2.16$ provides a useful comparison, hosting an extended network of satellites, diffuse gas, and filamentary Ly$\alpha$ emission over $\sim150$ kpc \citep{2006ApJ...650L..29M,2014A&A...570A..55D}. J0800+4029 differs in that it exhibits strong IHL and tidal features within the core, but lacks a similarly obvious extreme starburst component. While both systems are dynamically young, the observational emphasis in J0800+4029 is on disturbed redistribution rather than on a centrally dominant Ly$\alpha$ halo.

\subsection{Simulations and Timescales}

Recent models predict coevolution of Ly$\alpha$ emission and diffuse stellar components \citep{2025arXiv250320857K}. Hydrodynamical simulations such as IllustrisTNG, EAGLE, and Hydrangea consistently predict that diffuse stellar envelopes can emerge before the brightest cluster galaxy is fully assembled. The presence of a measurable diffuse-light component in J0800+4029 at $z\sim2$ is qualitatively consistent with that picture.

At the same time, the current observations do not yet permit a detailed timescale decomposition of stripping, cooling, and inflow. The spatial overlap of Ly$\alpha$ emission and IHL suggests that stellar and gaseous redistribution are occurring on comparable scales in this system, but a broader sample and a fully homogenized analysis will be needed to determine whether such behavior is generic.

\subsection{Implications for Cluster Assembly}

The coexistence of diffuse Ly$\alpha$ emission, a redshifted absorption-like feature, and intra-halo light reveals a multiphase medium characteristic of early cluster assembly. J0800+4029 provides a particularly useful case because the stellar debris and the diffuse gas can be studied jointly in a system that is clearly not dynamically relaxed.

If the apparent link between IHL-rich tidal structure and offset Ly$\alpha$ emission is confirmed in larger samples, it could become a useful diagnostic of baryonic redistribution during cluster growth. Future wide-field imaging and IFU spectroscopy will be important for testing whether the behavior seen in J0800+4029 is common among dynamically young protoclusters or whether it represents a more specialized evolutionary pathway.

\section{Summary}\label{sec:summary}

We have presented KCWI and {\it HST} observations of the CARLA protocluster J0800+4029 at \(z = 1.986 \pm 0.014\), providing a detailed joint view of diffuse stellar and gaseous components in a forming cluster core.

The most striking result is the spatial offset between the quasar and the dominant extended Ly$\alpha$ emission. In the current analysis, no comparably robust extended Ly$\alpha$ signal is recovered at the quasar position, whereas the detected diffuse emission is dominated by one main region associated with the tidal stellar structure seen in the \emph{HST} image.
Using a standard $1\arcsec$ aperture and a 7-channel window around systemic Ly$\alpha$, we estimate a conservative $3\sigma$ quasar-position limit of $8.16\times10^{-18}$~\SBcgs. For comparison with halo-luminosity measurements, a quasar-centered halo matched in projected area and spectral width to the detected offset/tidal complex would have a $3\sigma$ luminosity limit of $4.20\times10^{42}$ erg~s$^{-1}$, while the total detected Ly$\alpha$ luminosity of the offset/tidal complex is $1.42\times10^{43}$ erg~s$^{-1}$.

The Ly$\alpha$ halo is kinematically disturbed but remains close to systemic in its brightest parts. Typical FWHM values are $\sim300$--$500$~km~s$^{-1}$ in the brightest regions, with broader values appearing mainly in lower-surface-brightness or less securely sampled pixels.

Using a non-parametric analysis of the \emph{HST} imaging, we measure an IHL fraction of \(18.8\%\) for the tidal structure associated with the Ly$\alpha$ emission. \rev{Diffuse regions in the comparison sample do not show comparably clear Ly$\alpha$ counterparts in the current KCWI field and therefore provide a useful internal comparison.}

We also identify a redshifted absorption-like feature within the diffuse Ly$\alpha$ region, offset by \(\sim 1000 \, \mathrm{km\,s^{-1}}\) from systemic. This feature points to the presence of cold gas embedded in the disturbed structure, although its physical interpretation is not unique and may involve inflow, radiative-transfer effects, or other complex kinematic processes.

Taken together, these results support a picture in which diffuse stellar light and observable Ly$\alpha$ emission are linked to the tidally disturbed protocluster core rather than to the quasar itself. J0800+4029 therefore provides a useful case study of how stars and gas can be redistributed during the early assembly of a massive cluster environment.

Looking forward, the combined analysis of intracluster light and extended Ly$\alpha$ emission offers a promising diagnostic of baryonic physics in dense regions at cosmic noon. Wider samples observed with deep imaging and IFU spectroscopy will be required to determine how common this offset Ly$\alpha$--IHL configuration is and how it relates to protocluster dynamical state.

\section*{Data Availability}
The data underlying this article will be shared upon reasonable request to the corresponding author.

\begin{acknowledgements}
This study was supported by the LabEx UnivEarthS, ANR-10-LABX-0023 and ANR-18-IDEX-0001.
This research has made use of the Keck Observatory Archive (KOA), operated by the W. M. Keck Observatory and the NASA Exoplanet Science Institute (NExScI), under contract with the National Aeronautics and Space Administration.
\end{acknowledgements}

\bibliographystyle{aa}
\bibliography{refs}
\end{document}